\documentstyle[12pt]{article}

\begin{document}
\begin{center}
{\Large{\bf Electromagnetic self-energies of vector mesons and
electromagnetic mass anomaly of the massive Yang-Mills particles}}\\[4mm]
Dao-Neng Gao\\
{\small Center for Fundamental Physics,
University of Science and Technology of China\\
Hefei, Anhui, 230026, People's Republic of China}\\[2mm]
Mu-Lin Yan\\
{\small China Center of Advanced Science and Technology (World Lab)\\
P.O.Box 8730, Beijing, 100080, People's Republic of China\\
and\\
Center for Fundamental Physics,
University of Science and Technology of China\\
Hefei, Anhui, 230026, People's Republic of China\footnote{Mailing
address}}
\end{center}

\begin{abstract}
\noindent
A systematic method developed by the authors to evaluate the
one-loop electromagnetic self-energies of the low-lying mesons is extended
to the calculation of the vector sector including $\rho$, $\omega$, and
$\phi$-mesons. The theoretical result of $\rho^0-\rho^\pm$ electromagnetic  
mass difference is in agreement with the measurements. An
interesting effect
called as electromagnetic mass anomaly of the massive Yang-Mills particles
is further discussed. There is no new parameter in this study.
\end{abstract}

\newpage

\section{Introduction}
\par

The investigation on the electromagnetic (EM) masses of the low-lying
mesons has a long history, however, which is mainly on evaluating the EM
masses of pseudoscalar $\pi$ and $K$-mesons (see Ref. \cite{GLY} and
references therein).
In Ref. \cite{GLY}(hereafter referred as paper I), we have presented a
systematic
investigation on the EM mass splittings of the low-lying
mesons including $\pi$, $K$, $a_1$, $K_1$, and $K^*$ in the framework of
the U(3)$_L\times$U(3)$_R$ chiral theory of mesons \cite{Li95}. The
present paper is
the continuation of paper I. We are to calculate the EM
self-energies of the vector $\rho$, $\omega$, and $\phi$ mesons to
one-loop order and $O(\alpha_{\rm EM})$.

Chiral quark model is originated by Weinberg \cite{W79}, and developed by
Manohar and Georgi \cite{MG84}. In Ref. \cite{LYL91}, the vector meson(
$\omega-$meson) is firstly introduced into this model in order to 
study quark spin contents in chiral soliton model. The author of
Ref. \cite{Li95} further extended it to include the low-lying 
1$^-$(vector) and 1$^+$(axial-vector) mesons (called as
U(3)$_L\times$U(3)$_R$ chiral theory of
mesons). The U(3)$_L\times$U(3)$_R$ chiral theory of mesons has been
investigated extensively \cite{GLY,Li97,LGY98,Li98} and its theoretical
results agree well with the data. 

Chiral perturbation theory (ChPT), which is expanded in powers
of derivatives of the meson fields, is rigorous and phenomenologically
successful in describing the physics of the pseudoscalar mesons
at very low energies \cite{GL85}. In Ref. \cite{WY98},
starting from the U(3)$_L\times$U(3)$_R$ chiral 
theory of mesons, and by using path integral method to integrate out 
the vector and axial-vector resonances, the authors have derived the
chiral coupling constants of ChPT ($L_1$, $L_2$, $L_3$,
$L_9$, and $L_{10}$). The results are in good agreement with
the experimental values of the $L_i$ at $\mu=m_\rho$ in ChPT. 
Therefore, the QCD constraints discussed
in Ref. \cite{EGLPR} are met by this theory.

It has been pointed out in \cite{Li95} that vector meson dominance
(VMD)\cite{KLZ,Sak69} has been
introduced into the U(3)$_L\times$U(3)$_R$ chiral theory of mesons. This
means
that the electromagnetic interaction of the mesons has been well
established, which makes it possible to evaluate the EM self-energies of
the low-lying mesons systematically.

In paper I, the logarithmic divergences coming from meson-loop
diagrams which contribute to the EM mass splittings of $\pi$, $K$, $a_1$,
$K_1$, and $K^*$ mesons have been factorized by using the intrinsic
parameter $g$ of the theory (see paper I for details).
However, when this method is extended to
the cases of $\rho$, $\omega$, and $\phi$-mesons, the circumstances will
be complicated. The Feynman diagrams contributing to the EM self-energies
of $\rho$-mesons can be divided into three kinds, which have been shown in
Fig. 1. Fig. 1.1 is tree diagram, and it contributes the finite result;
Fig. 1.2  involves only logarithmic divergences, which could
be evaluated by using the same method in paper I. In the framework of
the U(3)$_L\times$U(3)$_R$ chiral theory of mesons, the explicit one-loop
Feynman diagrams which belong to Fig. 1.2 will be drawn in Figs. 8, 9,
and 10; It will
meet some difficulties in computing Fig. 1.3 because the quadratic or
higher 
order divergences will emerge in the Feynman integrations of these
loop diagrams. Therefore, 
it is unsuitable to factorize these divergences by using $g$ in which only
the logarithmic one is involved.
 
However, if watching these Feynman diagrams carefully, one will find that
the contribution from Fig. 1.3 could be ignored. Considering one-loop
correction to the $\rho-\gamma$ vertex (Fig. 2), and comparing Fig. 1.3
with Fig. 2, one will see that the contribution of Fig. 1.3 to EM
self-energies of $\rho$-mesons should vanish if the one-loop renormalized  
$\rho-\gamma$ vertex is used in Fig. 1.1.  

The expressions of VMD in the U(3)$_L\times$U(3)$_R$ chiral theory of
mesons have been derived in Ref.  
\cite{Li95}, which can be read as 
\begin{eqnarray}
\frac{e}{f_\rho}\{-\frac{1}{2}F^{\mu \nu}\rho^0_{\mu \nu}+A^\mu
J_\mu^\rho\},\\
\frac{e}{f_\omega}\{-\frac{1}{2}F^{\mu \nu}\omega_{\mu\nu}+A^\mu
J_\mu^\omega\},\\
\frac{e}{f_\phi}\{-\frac{1}{2}F^{\mu \nu}\phi_{\mu \nu}+A^\mu
J_\mu^\phi\},
\end{eqnarray}
where
\begin{eqnarray}
\frac{1}{f_\rho}=\frac{1}{2}g,\;\;\;\frac{1}{f_\omega}=\frac{1}{6}g,\;\;\;
\frac{1}{f_\phi}=-\frac{1}{3\sqrt{2}}g.
\end{eqnarray}
$J_\mu^\rho$, $J_\mu^\omega$, and $J_\mu^\phi$ are the corresponding
hadronic currents for $\rho^0$, $\omega$, and $\phi$ mesons respectively.

By employing the $\rho-\gamma$ vertex in eq. (1), the decay width of
$\rho\rightarrow e^+ e^-$ is 
\begin{equation}
\Gamma(\rho\rightarrow e^+ e^-)=\frac{4\pi\alpha_{\rm
EM}^2}{f_\rho^2}\frac{m_\rho}{3}.
\end{equation}
Here $\alpha_{\rm EM}=\frac{e^2}{4\pi}$. When $g$=0.39 \footnote{if
setting
$g$=0.39 instead of $g$=0.35,
we will get much better fit for the data than that in Ref. \cite{Li95},
and $g$=0.39 has been set
in paper I  and Refs. \cite{Li97,LGY98,Li98}.},  
the numerical result of $\Gamma(\rho\rightarrow e^+ e^-)$ is 6.53 keV,
which is in good agreement with the experimental data 6.77$\pm$0.32 keV
\cite{PDG98}.
It is expected that one-loop corrections to the $\rho-\gamma$
vertex are very small, and can be ignored. 

It is enough to evaluate the contributions from Fig. 1.1 and 1.2 to EM
self-energies of $\rho$-mesons. The effect of Fig. 1.3 is included
by renormalizing the $\rho-\gamma$ vertex. 
In fact, Fig. 1.3 is not the 1PI (one particle irreducible) diagram 
for the one-loop EM self-energies of $\rho$-meson. Thus, it is
straightforward to use the method developed in paper I to factorize the
logarithmic divergences from mesonic loop diagrams, then to get a finite 
result of $\rho^0-\rho^\pm$ EM mass
difference. The difficulties in calculating the EM self-energies of
$\omega$ and $\phi$-mesons can be circumvented similarly. 

It has been pointed out by Sakurai \cite{Sak69} that there are two ways of 
writing down VMD. In eqs. (1)-(3) (called as VMD1), the photon-meson
coupling have two approaches: the first one is the direct coupling of
photon
and mesons ($A^\mu J^i_\mu$, $i$=$\rho$, $\omega$, and $\phi$), the second
one is the interaction through the neutral vector mesons including
$\rho^0$, $\omega$,
and $\phi$.

The other representation of VMD (called as VMD2) \footnote{In Ref.
\cite{OPTW},
the two representations of VMD have been referred as VMD1 and VMD2.},
in which the photon-mesons coupling must be through the neutral vector
mesons, is expressed as follows 
\begin{eqnarray}
-\frac{e m_\rho^2}{f_\rho}\rho^0_\mu A^\mu+\frac{1}{2}(\frac{e
m_\rho}{f_\rho})^2 A_\mu A^\mu,\\
-\frac{e m_\omega^2}{f_\omega}\omega_\mu A^\mu+\frac{1}{2}(\frac{e
m_\omega}{f_\omega})^2 A_\mu A^\mu,\\
-\frac{e m_\phi^2}{f_\phi}\phi_\mu A^\mu+\frac{1}{2}(\frac{e
m_\phi}{f_\phi})^2 A_\mu A^\mu.
\end{eqnarray}

As proved by Kroll, Lee, and Zumino \cite{KLZ}, and by Sakurai
\cite{Sak69}, VMD1 and VMD2 are equivalent in
describing the electromagnetic interaction of the mesons. Fortunately, in
the framework of the U(3)$_L\times$U(3)$_R$ chiral theory of mesons and
employing VMD2, the loop diagram (Fig. 1.3) contributing to the EM
self-energies of
the vector mesons which contains quadratic or higher order divergences
will disappear automatically, and only the logarithmic divergence is
involved in the rest loop diagrams. Our calculations show that, without
considering Fig. 1.3,
contributions to EM self-energies of $\rho$-mesons by using VMD1 
are entirely equivalent to ones by using VMD2
directly. 

The purposes of the present paper are twofold:

1.~The investigation of EM self-energies of the vector mesons has been
concerned by particle physicists. Here, we are to extend the method used
in paper I to calculate the one-loop
EM self-energies of the low-lying mesons in the case of vector mesons
including $\rho$, $\omega$, and $\phi$-mesons, which makes the
investigation in paper I much more complete and systematic. 

2.~An interesting effect, EM masses
anomaly of the massive Yang-Mills particles, has been revealed in
Ref. \cite{YG98}, which states that the non-Abelian gauge structure
of the massive Yang-Mills particles makes
the EM mass of the neutral $K^*$(892) larger than that of the charged 
one. This claim will be re-examined in the cases of $\rho$ meson and
other massive Yang-Mills particles elaborately.  

The paper is organized as follows. In Sec. 2, 
the equivalence of VMD1 and VMD2 is discussed briefly. As examples,  
we show that, in the framework of the U(3)$_L\times$U(3)$_R$ chiral
theory of mesons, VMD1 and VMD2
could reduce the same results of the pion EM form factor and the EM mass
splitting of $\pi^\pm-\pi^0$, and the well-known result of
$m_{\pi^\pm}^2-m_{\pi^0}^2$ obtained by Das, Guralnik, Mathur, Low, and
Young \cite{DGMLY} is reproduced. Sec. 3, the
EM self-energies of the vector mesons including $\rho$, $\omega$,and
$\phi$ are evaluated, and the EM mass anomaly of the massive
Yang-Mills particles is discussed. Sec. 4, we give the summary of 
the results.

\section{VMD1 and VMD2}

VMD predates the standard model. It has not been derived from the
standard model directly, but nevertheless enjoys phenomenological supports
in describing hadronic electromagnetic interactions. In the
U(3)$_L\times$U(3)$_R$ chiral theory of mesons, the basic lagrangian of
the theory is (Hereafter we use the notations in paper I)
\begin{eqnarray}
\lefteqn{{\cal L}=\bar{\psi}(x)(i\gamma\cdot\partial+\gamma\cdot v
+e_0Q\gamma\cdot A+\gamma\cdot a\gamma_{5}
-mu(x))\psi(x)}\nonumber \\
& &+{1\over 2}m^{2}_{1}(\rho^{\mu}_{i}\rho_{\mu i}+
\omega^{\mu}\omega_{\mu}+a^{\mu}_{i}a_{\mu i}+f^{\mu}f_{\mu})\nonumber\\
& &+{1\over2}m^2_2(K_{\mu}^{*a}K^{*a\mu}+K_1^{\mu}K_{1\mu})\nonumber \\ 
& &+\frac{1}{2}m^2_3(\phi_{\mu} \phi^{\mu}+f_s^{\mu}f_{s\mu})
+{\cal L}_{\rm EM}   
\end{eqnarray}
where $u(x)=exp[i\gamma_{5} (\tau_{i}\pi_{i}+\lambda_a K^a+\eta
+\eta^{\prime})]$($i$=1,2,3 and $a$=4,5,6,7),
$a_{\mu}=\tau_{i}a^{i}_{\mu}+
\lambda_a K^a_{1\mu}+(\frac{2}{3} +\frac{1}{\sqrt{3}} \lambda_8)f_{\mu}+(
\frac{1}{3}-\frac{1}{\sqrt{3}} \lambda_8)f_{s\mu},
v_{\mu}=\tau_{i}\rho^{i}_{\mu}+\lambda_a K_{\mu}^{*a}+(\frac{2}{3}+
\frac{1}{\sqrt{3}}\lambda_8)\omega_{\mu}+(\frac{1}{3}- \frac{1}{\sqrt{3}}
\lambda_8)\phi_{\mu}$, $A_\mu$ is the photon field, $Q$ is the electric 
charge operator of $u$, $d$ and $s$ quarks, $\psi$ is
quark-fields, $m$ is a parameter related to the quark condensate, and
${\cal L}_{\rm EM}$ is the kinetic lagrangian of photon field. Thus,
the use of path integral method to integrate out the quark fields 
will reduce the effective lagrangian of the mesons. 

Since the photon field
$A_\mu$ has been introduced into eq. (9), it is very natural to deduce
the explicit expressions of VMD in the mesonic effective lagrangian, which
are just the eqs.(1)-(3). The full lagrangian which describes the
interaction of photon and the neutral vector mesons could be expressed as
follows,
\begin{equation}
{\cal L}_{\rm VMD1}=-\frac{1}{4}F_{\mu\nu}F^{\mu\nu}-
\frac{1}{4}\rho^0_{\mu\nu}\rho^{0\mu\nu}+\frac{1}{2}m_\rho^2\rho^0_\mu
\rho^{0\mu}+\rho^{0\mu} J_\mu^\rho+
\frac{e}{f_\rho}(-\frac{1}{2}F^{\mu
\nu}\rho^0_{\mu \nu}+A^\mu
J_\mu^\rho).
\end{equation}
For simplicity, here we only discuss the interactions between $\rho^0$ and 
the photon. VMD1 is manifestly gauge invariant for  the neutral
vector meson $\rho^0$ coupling with the conserved current in strong
interaction. As pointed out by Sakurai \cite{Sak69}, there is an
equivalent way of writing down a gauge invariant expression for hadronic
electromagnetic interaction, which is
\begin{equation}
{\cal L}_{\rm VMD2}=-\frac{1}{4}F_{\mu\nu}F^{\mu\nu}-
\frac{1}{4}\rho^0_{\mu\nu}\rho^{0\mu\nu}+\frac{1}{2}m_\rho^2\rho^0_\mu
\rho^{0\mu}+\rho^{0\mu} J_\mu^\rho-\frac{e m_\rho^2}{f_\rho}\rho^0_\mu
A^\mu+\frac{1}{2}(\frac{e m_\rho}{f_\rho})^2 A_\mu A^\mu.
\end{equation}
From VMD1 to VMD2, one can use the following transformation,
\begin{eqnarray}
\rho^0_\mu\rightarrow\rho^0_\mu-\frac{e}{f_\rho}A_\mu,\nonumber\\
A_\mu\rightarrow\sqrt{1-(\frac{e}{f_\rho})^2}~A_\mu,\\
e\rightarrow e~ \sqrt{1-(\frac{e}{f_\rho})^2}.\nonumber
\end{eqnarray}
It seems that the mass term of the photon field in eq. (11) violates  
gauge invariance of the electromagnetic interaction. However, the
$\rho\cdot A$ term in this equation contributes an imaginary mass term of
the photon, which exactly cancels the contribution from mass term of 
the photon. To illustrate this, one could calculate the modification of the
photon propagator using the new interaction lagrangian including the new mass
term of the photon (Fig. 3).

From Fig. 3, one will obtain
\begin{eqnarray}
iD(k^2)=\frac{-i}{k^2}+\frac{-i}{k^2}\frac{-i e m_\rho}{f_\rho}
\frac{-i}{k^2-m_\rho^2}\frac{-i e m_\rho^2}{f_\rho}\frac{-i}{k^2}+...\nonumber\\
+\frac{-i}{k^2}\frac{i e^2 m_\rho^2}{f_\rho^2}\frac{-i}{k^2}+...\nonumber\\
=\frac{-i}{k^2}+\frac{-i}{k^2}\frac{i e^2 m_\rho^2}{f_\rho^2}
\frac{k^2}{k^2-m_\rho^2}\frac{-i}{k^2}+...\nonumber
\end{eqnarray}
Using the operator identity
\begin{eqnarray*}
\frac{1}{A-B}=\frac{1}{A}+\frac{1}{A}B\frac{1}{B}
 +\frac{1}{A}B\frac{1}{A}B\frac{1}{A}+...,
\end{eqnarray*}
we have
\begin{eqnarray}
iD(k^2)=\frac{-i}{k^2-\frac{e^2 m_\rho^2}
{f_\rho^2}\frac{k^2}{k^2-m_\rho^2}}
\longrightarrow \frac{-i}{k^2}(1-\frac{e^2}{f_\rho^2})
\end{eqnarray}
for small $k^2$. Thus, we are simply left with a change in  EM
coupling constant $e$
$$ e^2\longrightarrow e^2 (1-\frac{e^2}{f_\rho^2}). $$

In the following, as an example, we show the equivalence of VMD1 and
VMD2 by
calculating pion EM form factor. In Ref. \cite{Li95}, the interaction
${\cal L}_{\rho\pi\pi}$ has been derived 
\begin{eqnarray}
{\cal L}_{\rho\pi\pi}=\frac{2}{g}\epsilon_{ijk}\rho_\mu^i\pi^j\partial^\mu
\pi^k-\frac{2}{\pi^2f_\pi^2 g}[(1-\frac{2c}{g})^2-4\pi^2
c^2]\epsilon_{ijk}\rho_\mu^i\partial_\nu\pi^j\partial^\mu\partial^\nu\pi^k,
\end{eqnarray}
where $c=\frac{f_\pi^2}{2 g m_\rho^2}$, and
$f_\pi$ is the decay constant of pion  with value $0.186$ GeV. 
Thus, the
pion EM form factor $F_\pi(k^2)$ can be defined through the amplitude
of process $\gamma\longrightarrow \pi^+\pi^-$
\begin{equation}
{\cal M}^\mu_{\gamma\rightarrow\pi^+\pi^-}=-e(q^+-q^-)^\mu F_\pi(k^2),
\end{equation}
here $q$ and $k$ are momentum of pion and the virtual photon respectively.
From VMD1 (as shown in Fig. 4.1), we have
\begin{equation}
F_\pi(k^2)=(1+\frac{k^2}{m_\rho^2-k^2})\frac{f_{\rho\pi\pi}(k^2)}{f_\rho},
\end{equation}
where
\begin{eqnarray*}
f_{\rho\pi\pi}(k^2)=\frac{2}{g}\{1+\frac{k^2}{2\pi^2
f_\pi^2}[(1-\frac{2c}{g})^2-4\pi^2 c^2]\}.
\end{eqnarray*}
Note that the second term in eq. (16) due to the direct interaction of
$\rho$ and $\gamma$ vanishes at $k^2=0$, and it is therefore very
important to
take into account the effect of the contact $J_\mu^\rho A^\mu$
interaction.

From VMD2 (as shown in Fig. 4.2), one obtains
\begin{equation}
F_\pi(k^2)=\frac{m_\rho^2}{m_\rho^2-k^2}\frac{f_{\rho\pi\pi}(k^2)}{f_\rho}.
\end{equation}
It is obvious that the two approaches lead to the identical result from
eq. (16) and eq. (17).  

Now, we show that VMD1 and VMD2 are equivalent in evaluating the EM mass 
splitting of pions. In paper I, the EM mass difference between $\pi^\pm$
and $\pi^0$ in the chiral limit has been obtained in terms of VMD1. Here
we will re-examine this calculation by employing  VMD2. 

The Feynman diagrams contributing to $(m_{\pi^\pm}^2-m_{\pi^0}^2)_{\rm
EM}$ have been shown in Fig. 5 and Fig. 6, which correspond to ones from
VMD1 and VMD2 respectively (The diagrams which receive contributions from
the interaction ${\cal L}_{\rho\pi\pi}$ have been neglected because their
contributions vanish in the chiral limit. This point has been demonstrated
in
paper I). Since there are no the direct vertices of $\pi$ and $\gamma$,
the number of the Feynman diagrams contributing to EM self-energies of
pions in terms of VMD2 has been much more decreased. 

Comparing Fig. 5 and Fig. 6, one will find that if the effective
propagators in Fig. 7.1 and Fig. 7.2 are equivalent, the contributions of
Fig. 5 and Fig. 6 must be the same to the EM self-energies of pion-mesons. 
Note that in Fig. 7.1, the $\rho-\gamma$ vertex comes from VMD1, which is
momentum dependent; in Fig. 7.2, the $\rho-\gamma$ vertex is from VMD2. 
It is easy to prove that Fig. 7.1 and Fig. 7.2 both contribute 
\begin{equation}
\frac{-i}{k^2}\frac{e^2}{f_\rho^2}[\frac{m_\rho^4}{(k^2-m_\rho^2)^2}
(g_{\mu\nu}-\frac{k_\mu k_\nu}{k^2})+a\frac{k_\mu k_\nu}{k^2}],
\end{equation}
where $a$ is the gauge parameter. Because there is always a factor
$\frac{e}{f_\rho}$ in the coupling of the
photon with the current $J_\mu^\rho$ in VMD1, this factor has been
considered in calculating Fig. 7.1. Eq. (18) has been obtained by 
Lee and Nieh in Ref. \cite{LN68}. Therefore, it is not surprising that the
result of EM self-energies of pions in terms of VMD1 is the same as one in
terms of VMD2. 

The straightforward calculation of Fig. 6 leads to 
\begin{equation}
(m_{\pi^\pm}^2-m_{\pi^0}^2)_{\rm EM} =i{e^2 \over f_\pi^2}
{\int \frac{d^4 k}{(2\pi)^4} (D-1)m_\rho^4} {{(F^2+{k^2 \over 2\pi^2})}
\over
{k^2 (k^2-m_\rho^2)^2 }}
[1+{\gamma^2 \over g^2}{{F^2+{k^2 \over 2\pi^2}} \over
{k^2 -m_a^2} }].
\end{equation}
This is just eq. (34) in paper I, which is derived in terms of VMD1.
The gauge independence of $(m_{\pi^\pm}^2-m_{\pi^0}^2)_{\rm EM}$
 can be
proven in the same way as that in paper I.

In deriving eq. (19) or eq. (34) in paper I,
the calculations on EM-masses are up to the fourth order
covariant derivatives in effective lagrangians \cite{Li95}.
In the remainder of this section,
we find that the well-known result of
$(m_{\pi^\pm}^2-m_{\pi^0}^2)_{\rm EM}$
given by
Das et al \cite{DGMLY} can be reproduced if the EM
self-energies of 
pions receives the contributions only from the second order derivative
terms.
In this case, the interaction lagrangians ${\cal L}_{\rho\rho\pi\pi}$ and
${\cal L}_{\rho\pi a}$ [eqs.(18)(19) in paper I] will be simplified as
follows
\begin{eqnarray*}
& &{\cal L}_{\rho\rho\pi\pi}=\frac{2F^2}{g^2 f_\pi^2}\rho^i_\mu
\rho^{j\mu}(\pi^2\delta_{ij}-\pi_i\pi_j),\;\;\;\;\;
{\cal L}_{\rho\pi a}=-\frac{2F^2}{f_\pi
g^2}\rho^i_{\mu}\epsilon_{ijk}\pi^k
a^{j\mu}.
\end{eqnarray*}
Thus, in the chiral limit, the EM mass difference of pions is
\begin{eqnarray}
(m_{\pi^\pm}^2-m_{\pi^0}^2)_{\rm EM}
=\frac{i3e^2}{f_\pi^2}
\int\frac{d^4k}{(2\pi)^4}\frac{m_\rho^4 F^2}{k^2(k^2-m_\rho^2)^2}
(1+\frac{F^2}{g^2(k^2-m_a^2)}).
\end{eqnarray}
The Feynman integration in eq.(20) is finite. So it is
straightforward to get the result of $(m_{\pi^\pm}^2-m_{\pi^0}^2)_{\rm 
EM}$
after performing this integration, which is
\begin{equation}
(m_{\pi^\pm}^2-m_{\pi^0}^2)_{\rm EM}
=\frac{3\alpha_{\rm EM}m_\rho^4}{8\pi f_\pi^2}
\{\frac{2F^2}{m_\rho^2}-\frac{2F^4}{g^2(m_a^2-m_\rho^2)}(\frac{1}{m_\rho^2}
+\frac{1}{m_a^2-m_\rho^2}{\rm log}{\frac{m_\rho^2}{m_a^2}})\}.
\end{equation}
Because we only consider the
second order derivative terms in the
lagrangian, the relation between $m_a$ and $m_\rho$ is
$m_a^2=\frac{F^2}{g^2}
+m_\rho^2$ instead of eq.(4) in paper I.
Thus, we can get
\begin{equation}
(m_{\pi^\pm}^2-m_{\pi^0}^2)_{\rm EM}
=\frac{3\alpha_{\rm EM}}{4\pi}\frac{m_a^2 m_\rho^2}{m_a^2-m_\rho^2}
{\rm log}{\frac{m_a^2}{m_\rho^2}}.
\end{equation}
When substituting the relation $m_a^2=2 m_\rho^2$, which can be derived
from
the Weinberg sum rules \cite{Wb67}, into eq.(21), we have
\begin{equation}
(m_{\pi^\pm}^2-m_{\pi^0}^2)_{\rm EM}
=\frac{3{\rm log}2}{2\pi}\alpha_{\rm EM} m_\rho^2,
\end{equation}  
which is exactly the result obtained by Das et al. \cite{DGMLY}, and
serves as the leading term of eq.(19).

\section{EM self-energies of the vector mesons and EM mass
anomaly of the massive Yang-Mills particles}

The subject of EM contributions of vector mesons is in fact very old. The
early attempts are in \cite{LN68,S64,C69}, and the recent are in
\cite{BG96,AA99}.
In Ref. \cite{GLY,YG98}, the EM masses of $K^*(892)$ have been evaluated
in the framework of the U(3)$_L\times$U(3)$_R$ chiral theory of mesons. In
this section, we will deal with the EM self-energies of $\rho$,
$\omega$, and $\phi$-mesons to
one-loop order and $O(\alpha_{\rm EM})$. 

The one-loop Feynman diagrams contributing to the EM self-energies of
$\rho$-meson are only from the massive Yang-Mills
(MYM)self-interactions 
and Gauging Wess-Zumino-Witten (GWZW) anomaly interactions in the
effective lagrangian of the U(3)$_L\times$U(3)$_R$ chiral theory of mesons
\cite{Li95}.  This case is the same as that of $K^*(892)$-meson
\cite{YG98}.

From Ref. \cite{Li95}, MYM lagrangian related to $\rho$-meson is
\begin{equation}
{\cal L}_{\rm MYM}=-{1 \over 8} Tr (\partial_\mu \rho_\nu-\partial_\nu
\rho_\mu-
\frac{i}{g}[\rho_\mu, \rho_\nu]-\frac{i}{g}[a_\mu, a_\nu])^2+ {\rm
mass\; terms\; of }\; \rho.
\end{equation}
Thus, the three-point and four-point MYM interaction vertices which
contribute to EM self-energies of $\rho$-meson 
can be read from eq. (24)
\begin{eqnarray}
{\cal L}_{\rm MYM}^{(3)}=\frac{1}{g}\epsilon_{ijk}
\partial_\mu\rho^i_\nu\rho^{j\mu}\rho^{k\nu},\\
{\cal L}_{\rm MYM}^{(4)}=\frac{1}{g^2}(\rho^i_\mu\rho^i_\nu
\rho^{j\mu}\rho^{j\nu}-\rho^i_\mu \rho^{i\mu}\rho^j_\nu\rho^{j\nu}).
\end{eqnarray}

The GWZW anomaly lagrangian in the present theory contributing to EM
self-energies  of $\rho$-meson is \cite{Li95} 
\begin{equation}
{\cal L}_{\rm GWZW}^{\omega\rho\pi}=-\frac{3}{\pi^2 g^2 f_\pi}
\epsilon^{\mu\nu\alpha
\beta}\partial_\mu\omega_\nu\rho_\alpha^i \partial_\beta\pi^i.
\end{equation}

Now, combining VMD1 [eqs.(1),(2)] or VMD2 [eqs. (6), (7)], one can
evaluate the one-loop EM self-energies of $\rho$-meson in the standard
way  presented in paper I. The corresponding diagrams are shown in Figs.
8, 9, and 10.  Note that all
the diagrams will contribute to EM self-energies of $\rho$-meson if
we use VMD1 while only Figs. 8.3, 9.3 and 10.3 give contributions if
using VMD2. In the following, we calculate these diagrams separately.

From Fig. 8, the EM mass square difference corresponding to the
contributions of three-point MYM interactions can be calculated
directly, which is 
\begin{eqnarray}
& &m^2_{\rm EM}(\rho^\pm)_{\rm MYM}^{(3)}=
i e^2 \int\frac{d^4 k}{(2\pi)^4}\frac{m_\rho^4}{k^2(k^2-2 p\cdot
k)(k^2-m_\rho^2)^2}\nonumber\\
& &\times [3k^2+4 m_\rho^2-\frac{(k^2)^2}{m_\rho^2}-4\frac{(k\cdot
p)^2}{k^2}+2k\cdot p
+\frac{\langle(k\cdot\rho^{\underline{i}})(k\cdot\rho^{\underline{i}})
\rangle}{\langle~\rho^{\underline{i}}\rho^{\underline{i}}~\rangle}
(9+\frac{k^2}{m_\rho^2}-\frac{2k\cdot p}{k^2})],
\end{eqnarray}
where $p$ is 4-momentum of $\rho$-meson, and $p^2=m_\rho^2$;
$\langle~\rho^{\underline{i}}\rho^{\underline{i}}~\rangle=\int d^4 x
\langle~\rho|\rho_\mu^{\underline{i}}(x)\rho^{\underline{i}\mu}(x)|\rho~\rangle$,
and $\underline{i}=1,~2$. After performing the Feynman integrations 
in eq. (28), we have
\begin{equation}
m^2_{\rm EM}(\rho^\pm)_{\rm MYM}^{(3)}=-e^2 m_\rho^2
[-\frac{3}{4}
\chi_\rho+\frac{1}{16\pi^2}(\frac{11\pi}{4\sqrt{3}}-\frac{37}{12})].   
\end{equation}
The logarithmic divergence is involved in $\chi_\rho$, which has been
factorized by using the intrinsic parameter $g$ of the theory in
paper I,
\begin{eqnarray}
\chi_\rho=\frac{1}{g^2}+\frac{1}{32\pi^2}+\frac{1}{16\pi^2}{\rm
log}\frac{f_\pi^2}{6(g^2 m_\rho^2-f_\pi^2)}.
\end{eqnarray}
$g=0.39$ has been fixed in paper I. Substituting the experimental
values of $f_\pi$ and $m_\rho$ into eq. (29), we obtain the numerical
result 
\begin{equation}
m^2_{\rm EM}(\rho^\pm)_{\rm MYM}^{(3)}=-3.36\times
10^{-4}~{\rm GeV^2}.
\end{equation}

Similarly, the contribution to EM self-energies of $\rho$-meson (Fig. 9)
from four-point MYM interaction is
\begin{equation}
m^2_{\rm EM}(\rho^\pm)_{\rm MYM}^{(4)}=-ie^2
\frac{9m_\rho^4}{4}\int \frac{d^4
k}{(2\pi)^4}\frac{1}{k^2(k^2-m_\rho^2)^2}.
\end{equation}
It is remarkable that eq. (32) is free of the divergence and 
independent of $g$,
so it is straightforward to get 
\begin{eqnarray}
m^2_{\rm EM}(\rho^\pm)_{\rm MYM}^{(4)}=-\frac{9e^2}{64 \pi^2 m_\rho^2}.
\end{eqnarray}
Numerically
\begin{equation}
m^2_{\rm EM}(\rho^\pm)_{\rm MYM}^{(4)}=-7.75\times
10^{-4} ~{\rm GeV^2}.
\end{equation}

The contribution to the EM self-energies of $\rho$-mesons from GWZW
anomaly can be evaluated by using eq. (27) [Fig. 10]. However, it is easy
to see that the contribution of Fig. 10 leads to the same shift of
the charged and neutral $\rho$-meson masses, so it has no contribution to
the EM mass difference of $\rho^\pm-\rho^0$. The EM self-energies of
$\rho$-meson in this case is easily obtained
\begin{eqnarray}
m^2_{\rm EM}(\rho^0)_{\rm GWZW}=m^2_{\rm EM}(\rho^\pm)_{\rm GWZW}=
\frac{e^2 m_\rho^4}{128 \pi^6 g^2 f_\pi^2}
=9.96\times 10^{-5}~{\rm GeV^2}.
\end{eqnarray}
There is also no divergences involved in Fig. 10.  We take
$m_\rho^2=m_\omega^2$ in deriving eq. (35), and its numerical result is
small. 

Different from the case of $K^*(892)$-meson, the direct $\rho^0$-photon
coupling which comes from VMD can bring the tree diagram contributing to
EM masses of $\rho^0$, which has been shown in Fig. 1.1. It is easy to
evaluate its contribution 
\begin{equation}
m^2_{\rm EM}(\rho^0)_{\rm Tree}=-\frac{e^2 g^2}{4}m_\rho^2.
\end{equation}
Numerically
\begin{equation}
m^2_{\rm EM}(\rho^0)_{\rm Tree}=-2.06\times 10^{-3}~{\rm GeV^2}.
\end{equation}

Totally, $\rho^0-\rho^\pm$ EM mass difference is 
\begin{eqnarray}
(m_{\rho^0}^2-m_{\rho^\pm}^2)_{\rm EM}=-9.49\times 10^{-4}~{\rm GeV^2},\\
(m_{\rho^0}-m_{\rho^\pm})_{\rm EM}=-0.62 ~{\rm MeV}.
\end{eqnarray}

The light quark mass terms of the QCD lagrangian are 
$$
{\cal L}_m=-m_u \bar{u} u-m_d \bar{d} d -m_s \bar{s}s- ...
$$
It is important to note that the isospin violating piece of ${\cal L}_m$,
 $-\frac{1}{2}(m_d-m_u)(\bar{d} d-\bar{u} u)$, transforms as $\Delta I=1$
under the isospin subgroup. 
The operator which produces the $\pi^\pm-\pi^0$ (or $\rho^\pm-\rho^0$) 
mass difference must have $\Delta I=2$, therefore
$(m_{\pi^\pm}-m_{\pi^0})_{\rm qm}=
(m_{\rho^\pm}-m_{\rho^0})_{\rm qm}=0$  at the leading order in quark mass
expansion (The subscript qm denotes the contribution from quark mass).
This has been verified in the
case of pions, as is well known that the $\pi^\pm-\pi^0$ mass difference
is almost entirely electromagnetic in origin. 

The experimental value of $m_{\rho^0}-m_{\rho^\pm}$ \cite{PDG98}
has a large error bar
\begin{equation}
m_{\rho^0}-m_{\rho^\pm}=0.1\pm0.9 ~{\rm MeV}.
\end{equation}
Recent result obtained by ALEPH Collaboration \cite{ALEPH97} is
\begin{equation}
m_{\rho^0}-m_{\rho^\pm}=0.0\pm1.0 ~{\rm MeV}.
\end{equation}
Comparing with eqs. (40) and (41), one will find that our prediction [eq.
(39)] is in agreement with the measurements. This means that EM mass
correction to the mass difference of $\rho$-meson is
important, although its value is much smaller than the corresponding one
of pions.

The EM self-energies of $\omega$-meson can also be calculated. The
tree level contribution from $\omega\longrightarrow\gamma\longrightarrow 
\omega$, which is 
\begin{equation}
m^2_{\rm EM}(\omega)_{\rm Tree}=-\frac{e^2 g^2
m_\omega^2}{36}=-2.37\times 10^{-4}~{\rm GeV^2}.
\end{equation}
Besides the contribution from tree diagram, eq. (27) also contributes
one-loop diagrams to the EM self-energies of $\omega$-meson (as shown
in Fig. 11), and the loop diagrams give the finite contributions,
\begin{equation}
m^2_{\rm EM}(\omega)_{\rm 1-LOOP}=\frac{9e^2m_\rho^2m_\omega^2}{128 \pi^6
g^2 f_\pi^2}=4.62\times 10^{-4}~{\rm
GeV^2}.
\end{equation}
Thus the numerical result of the total EM self-energies of $\omega$-meson
to one-loop order and
$O(\alpha_{\rm EM})$ is 
\begin{eqnarray}
m^2_{\rm EM}(\omega)_{\rm total}=2.25\times 10^{-4}~{\rm GeV^2},\\
(m_\omega)_{\rm EM}=0.14~{\rm MeV}.
\end{eqnarray} 

In the framework of the U(3)$_L\times$U(3)$_R$ chiral theory of mesons,
the vector mesons are written as 
\begin{eqnarray*}
v_{\mu}=\tau_{i}\rho^{i}_{\mu}+\lambda_a K_{\mu}^{*a}+(\frac{2}{3}+
\frac{1}{\sqrt{3}}\lambda_8)\omega_{\mu}+(\frac{1}{3}-
\frac{1}{\sqrt{3}}\lambda_8)\phi_{\mu},
\end{eqnarray*}
which means that the $\omega$-meson is free of $s\bar{s}$, $\phi$-meson
is pure $s\bar{s}$, and the mixing of $\omega$ and $\phi$-meson is ideal.
This is implied by the Okubo-Zweig-Iizuka (OZI) rule \cite{OZI}. In
the large $N_C$ limit, the
calculation of the EM self-energies of $\phi$-meson will be much simpler,
only the tree level contribution from $\phi\longrightarrow
\gamma\longrightarrow \phi$. Therefore we have
\begin{eqnarray}
m^2_{\rm EM}(\phi)=-\frac{e^2 g^2 m_\phi^2}{18}=-8.06\times 10^{-4}~{\rm
GeV^2},\\
(m_\phi)_{\rm EM}=-0.40 ~{\rm MeV}.
\end{eqnarray}

Empirically, the U(3)$_L\times$U(3)$_R$ symmetry is broken due to the
strong U(1) anomaly. Therefore the corrections from the next leading order
corrections of large
$N_C$ expansion to the processes which are related to the $\omega$ and
$\phi$ mixing are non-trivial. However, this is beyond the scope of the
present paper.

In Ref. \cite{YG98}, the EM masses of $K^*(892)$-meson has been
calculated, especially, one effect called as 
EM mass anomaly of massive Yang-Mills particles has been revealed. 
In the
case of $\rho$-meson, it is easy to find that this conclusion also
hold
from eqs. (31) and (34) because the three-point and four-point MYM
interactions contribute the negative values to the EM masses of
$\rho^\pm$.
Actually, this effect can also be seen in evaluating the EM self-energies
of the axial-vector mesons $a_1$ and $K_1$ for these mesons are
also introduced
into the U(3)$_L\times$U(3)$_R$ chiral theory of mesons as the massive
Yang-Mills particles. In paper I, EM masses of $a_1$ and $K_1$ have been
calculated. Here we extract the contributions coming from the interaction
lagrangians related to the three-point and four-point MYM vertices,  
\begin{eqnarray*}
m_{\rm EM}^2(a^\pm)^{(3)}_{\rm MYM}=-0.002688~{\rm GeV^2},\\
m_{\rm EM}^2(a^\pm)^{(4)}_{\rm MYM}=-0.000648~{\rm GeV^2},
\end{eqnarray*}
for $a_1$-meson, and
\begin{eqnarray*}
& &[m_{\rm EM}^2(K_1^\pm)-m_{\rm EM}^2(K_1^0)]^{(3)}_{\rm MYM}
=-0.003474~{\rm GeV^2},\\
& &[m_{\rm EM}^2(K_1^\pm)-m_{\rm EM}^2(K_1^0)]^{(4)}_{\rm MYM}
=-0.000781~{\rm GeV^2},
\end{eqnarray*}
for $K_1$ meson. Here, it is obvious that the contributions from the MYM
self-interaction make the EM masses of the neutral mesons larger than that
of the charged one for the massive Yang-Mills particles.  
  
It is expected that this is a non-trivial and unusual effect for the
massive Yang-Mills particles because it is contrary to the common
knowledge of the hadron's EM-masses, such as  $m({\rm
neutron})_{\rm EM}<m({\rm proton})_{\rm EM}$,
$m(\pi^0)_{\rm EM}<m(\pi^+)_{\rm EM}$, and $m(K^0)_{\rm
EM}<m(K^+)_{\rm EM}$ \cite{GL82}. The studies in Refs. \cite{YG98,GY98}
show that the experiment favors to support this EM masses anomaly effect.
However, the experimental information for the EM masses of the vector and
axial-vector mesons is rather poor, so this effect needs to be further
investigated.

\section{Summary}

In terms of two equivalent representations of VMD (VMD1 and VMD2), the
method developed in paper I to calculate the one-loop EM self-energies
of the low-lying mesons has been naturally extended to the case of the
vector sector including $\rho$, $\omega$, and $\phi$-mesons.  
Because all the parameters of the theory have been fixed previously, this
is a parameter free study in the present paper.

The theoretical result of the EM mass difference of $\rho$ meson is in
agreement with the measurements, which means that the EM contribution is
important to the $\rho^0-\rho^\pm$ mass difference. 
The EM self-energies of $\omega$ and $\phi$-mesons which make the
shifts for the mass of the mesons are also evaluated. 

It has been shown that the effect of EM mass anomaly of the massive
Yang-Mills particles, which has been revealed in Ref. \cite{YG98}, holds
in the case of the $\rho$-meson. An elaborate analysis of the cases of
$a_1$ and $K_1$ also supports this conclusion. This interesting effect
should be further investigated, and also be tested by the experiments in
the future.

\begin{center} {\bf ACKNOWLEDGMENTS} \end{center}   
This work is partially supported by NSF of China
through Chen Ning Yang and the Grant LWTZ--1298 of Chinese Academy of
Sciences.

\vskip1.0cm
\leftline{\bf Caption}
\begin{description}
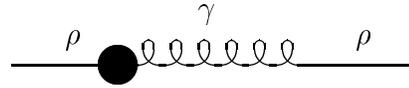
\item[Fig. 1] The diagrams contributing to electromagnetic self-energies
of $\rho-$meson up to one-loop order.
\item[Fig. 2] The $\rho-\gamma$ vertex and its one-loop corrections.
\item[Fig. 3] The modification of the photon propagator by using VMD2. The
{\bf $\times$} denotes the contribution from the mass term of the
photon in eq. (11).
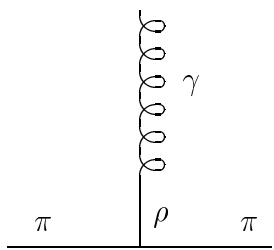
\item[Fig. 4] The pion electromagnetic form factor in terms of VMD1 and
VMD2. Fig. 4.1 is from VMD1, and Fig. 4.2 is from VMD2.
\item[Fig. 5] The one-loop diagrams contributing to electromagnetic mass
difference between $\pi^\pm$ and $\pi^0$ by using VMD1.
\item[Fig. 6] The one-loop diagrams contributing to electromagnetic mass  
difference between $\pi^\pm$ and $\pi^0$ by using VMD2.
\item[Fig. 7] The effective propagators coming from VMD. Fig. 7.1 is in
terms of VMD1, Fig. 7.2 is in terms of VMD2.
\item[Fig. 8] The one-loop diagrams contributing toelectromagnetic
self-energies of $\rho-$meson from three-point MYM interactions. All
three diagrams give contributions if using VMD1 while only Fig. 8.3 does  
if using VMD2.
\item[Fig. 9] The one-loop diagrams contributing to electromagnetic
self-energies of $\rho-$meson from four-point MYM interactions. All
three diagrams give contributions if using VMD1 while only Fig. 9.3 does
if using VMD2.
\item[Fig. 10] The one-loop diagrams contributing to electromagnetic
self-energies of $\rho-$meson from GWZW anomaly lagrangians.
All three diagrams give contributions if using VMD1 while only Fig. 10.3
does if using VMD2.
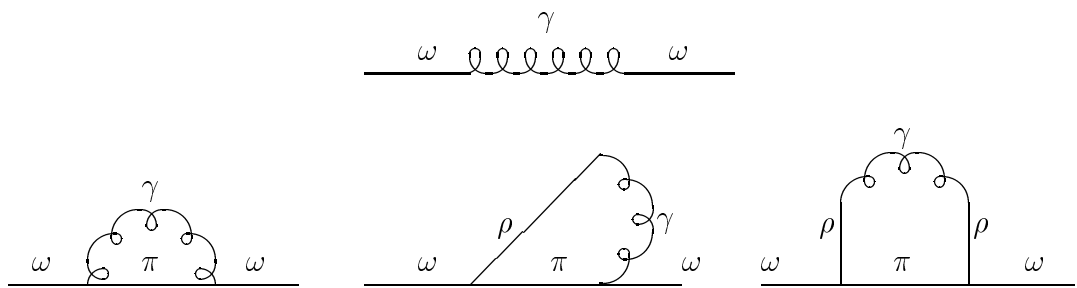
\item[Fig. 11] The tree and one-loop diagrams contributing to
electromagnetic self-energies of $\omega-$meson.
\end{description}

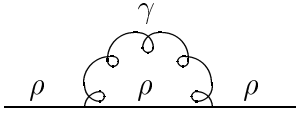
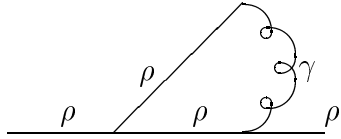
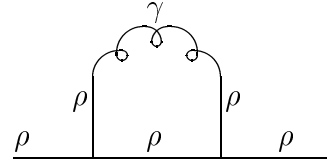
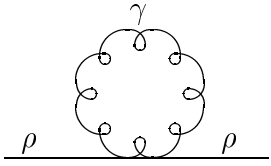
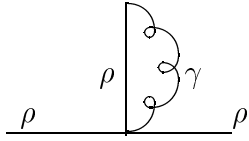
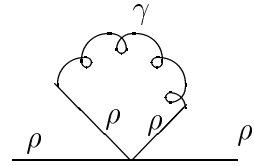
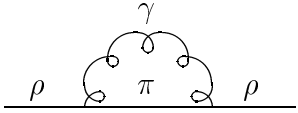
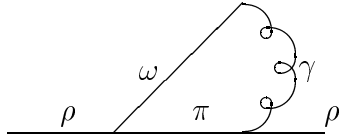
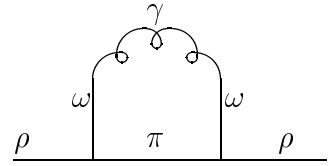
\end{document}